\documentclass{elsart}

\usepackage{graphicx}
\usepackage{amssymb}
\newcommand{\vsd}{V}
\newcommand{\Vsd}{$\vsd$}
\newcommand{\GVB}{$G\left(\vsd,B\right)$}
\newcommand{\micro}[1]{$\mu\mbox{#1}$}
\newcommand{\fig}[1]{Fig.~\ref{#1}}
\newcommand{\figs}[1]{Figs.~\ref{#1}}
\newcommand{\eq}[1]{Eq.~\ref{#1}}

\newcommand{\text}[1]{\mbox{\tiny #1}}
\begin{document}

\begin{frontmatter}



\title{Many-body dispersions in interacting ballistic quantum wires}

\author{Ophir M. Auslaender, Hadar Steinberg and Amir Yacoby}
\address{Department of Condensed Matter Physics, Weizmann Institute of Science, Rehovot 76100, Israel}
\author{Yaroslav Tserkovnyak and Bertrand I. Halperin}
\address{Lyman Laboratory of Physics, Harvard University,
Cambridge, Massachusetts 02138, USA}%
\author{Rafael de Picciotto, Kirk W. Baldwin, Loren N. Pfeiffer}\newline\author{and Ken W. West}
\address{Bell Labs, Lucent Technologies, 700 Mountain Avenue, Murray Hill, New Jersey 07974, USA}

\begin{abstract}
We have measured the collective excitation spectrum of interacting electrons in
one-dimension. The experiment consists of controlling the energy and momentum
of electrons tunneling between two clean and closely situated, parallel quantum wires
in a GaAs/AlGaAs heterostructure while measuring the resulting conductance. We
measure excitation spectra that clearly deviate from the non-interacting spectrum,
attesting to the importance of Coulomb interactions. Notable is an observed 30\% enhancement of the velocity of the main excitation branch relative to non-interacting
electrons with the same density. In short
wires, finite size effects resulting from broken translational invariance are observed. Spin - charge separation is manifested through moir\'{e} patterns, reflecting different spin and charge excitation velocities.
\end{abstract}

\begin{keyword}
A. nanostructures\sep D. electron-electron interactions\sep D. electronic transport\sep D. tunnelling
\PACS 73.21.Hb\sep71.10.Pm\sep73.23.Ad\sep73.50.Jt
\end{keyword}
\end{frontmatter}
\newpage
The problem of interacting electrons in one dimension has attracted considerable attention over the years, both theoretically and experimentally. One dimensional systems differ fundamentally from their higher dimensional counterparts in the role interactions play. The latter are well described by Landau Fermi-liquid theory \cite{Nozieres97}, in which a system of interacting electrons is mapped onto a system of weakly interacting long-lived quasiparticles \cite{Altshuler85}, each carrying a fundamental unit of charge, $-e$, and a fundamental unit of spin, $\hbar/2$. In one dimension the quasiparticles become unstable because of their large probability to decay into particle-hole pairs, a process that is blocked in higher dimensions due to phase-space constraints. A 1D system is best described as a Luttinger-liquid \cite{Emery79,Haldane81}, characterized by decoupled long-range spin and density correlations. It is predicted that an injected electron will break up into separate spin and charge excitations which propagate with separate velocities and give separate singularities in the one-electron Green function.

In the experimental study of clean one-dimensional (1D) systems we search for effects that are a direct and exclusive consequence of interactions and the correlations they induce. The first transport studies of wires aimed at measuring the conductance $G=\partial I/\partial V_{SD}$ ($I$ is the current along a one-dimensional system subjected to a bias $V_{SD}$) \cite{Tarucha95,Yacoby96}. Only later was it realized that in clean, finite wires $G$ is quantized in units of $e^2/h$ regardless of the strength of interactions \cite{Maslov95a,Oreg96b}. The next step was the realization that disorder allows to tap into the unique properties of a Luttinger-liquid. In the limit of low energies, a single, weak scatterer in a 1D wire with repulsive interactions effectively cuts it into two disjoint pieces. Thus  $G$ reflects the long range correlations in a Luttinger-liquid through a power-law dependence on energy \cite{Kane92a,Kane92b,Kane93,Ophir00,Postma01}, that is absent in a Fermi-liquid.

When looking for ways to study a many body system the most fundamental quantity to focus on is its excitation spectrum. Naturally one would especially like to be able to measure it directly. First steps in this direction were experiments that used either Raman spectroscopy \cite{Goni91} or angle resolved photoemission \cite{Kim96,Segovia99,Claessen02,Lorenz02} to measure the excitation spectrum of electrons in one dimension. Below we show that measuring tunneling between two parallel wires enables to circumvent some of the limitations inherent in optical spectroscopy \cite{Altland99,Ophir02} and allows to map out the dispersions of electrons in one dimension up to and beyond the Fermi momentum with high accuracy. 

The main idea behind using tunneling amongst parallel wires for spectroscopy is that it affords simultaneous control of both the energy and the momentum transferred between the wires by the tunneling electrons. The energy is given by $eV$, where \Vsd\ is the bias voltage difference between the wires while the momentum is $\hbar q_B\equiv eBd$, where $B$ is a magnetic field applied perpendicular to the plane of the wires and $d$ is the distance between them. For long junctions the tunneling process conserves both energy and momentum, so unless there exists a state in the target wire at the particular value of energy and momentum selected by $V$ and $B$, tunneling is blocked. This allows to directly determine the dispersions of the many-body states (see e.g. \cite{Carpentier02,Zuelicke02}).

\begin{figure}
\centering\includegraphics[height=5cm,angle=0,clip=]{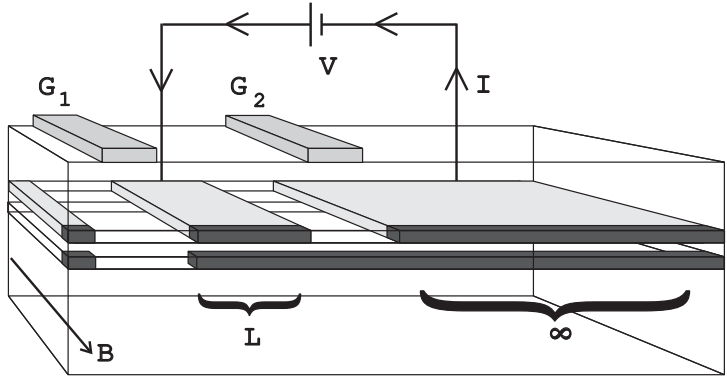}
\caption{\label{circuit}Schematic of the sample and the contacting scheme. The sample is fabricated using CEO. Two parallel 1D wires (dark gray) span along the whole cleaved
edge. The upper wire overlaps the 2DEG (light gray) in the upper quantum well (20nm thick), while the lower wire is separated from them by a 6nm insulating barrier in an otherwise empty quantum well (30nm thick). Contacts to the wires are made through the 2DEG. Several 2\micro{m}-wide tungsten top gates can be biased to deplete the electrons under them (only $G_1$ and $G_2$ are shown). The magnetic field $B$ is perpendicular to the plane defined by the wires. The depicted configuration allows to study the conductance of a single wire-wire tunnel junction of length $L$ by measuring the current $I$ that flows when a bias voltage $V$ is applied between the wires.}
\end{figure}%
The first study of tunneling between two parallel wires was facilitated by cleaved-edge overgrowth \cite{Pfeiffer93,Yacoby96,Yacoby97} of a AlGaAs/GaAs double quantum well heterostructure (see \fig{circuit}). An initial growth sequence renders the 20nm-wide upper quantum well occupied by a two-dimensional electron gas (2DEG) and the 30nm-wide lower quantum well devoid of electrons. The two quantum wells are separated by a 6nm-wide insulating AlGaAs barrier. After a 30sec infra-red illumination the mobility of the 2DEG is $\mu\approx3\times10^{6}\mbox{cm}^2\mbox{V}^{-1}\mbox{s}^{-1}$ and its density is $n\approx2\times10^{11}\mbox{cm}^{-2}$. After depositing 2\micro{m}\ tungsten gates on the top surface of the sample, it is reinserted into the molecular-beam epitaxy (MBE) chamber. The sample is then cleaved in this pristine environment to expose a clean (110) plane and a second growth sequence is initiated. As a result a trapping potential for electrons is created along the cleaved edge in each of the wells. The quantum states in this potential constitute the sub-bands in each of the wires in the experiment. Typically there are 3-5 occupied sub-bands in each of the wires following the 30sec illumination mentioned above. 

The measurements reported here were conducted in a ${}^3$He refrigerator at a base temperature of 0.25K (unless otherwise stated) using a lockin amplifier at a frequency of 14Hz with an excitation of 10\micro{V}. All of the measurements were two-terminal measurements between indium contacts to the 2DEG in the upper quantum well (see \fig{circuit}). The source is the 2DEG between gates $G_1$ and $G_2$. The bias on $G_1$ is set to deplete both wires and the 2DEG, while the bias on $G_2$ is set to leave only the lower wire conducting. Thus, the upper wire between $G_1$ and $G_2$ is at electrochemical equilibrium with the source 2DEG, while the whole semi-infinite lower wire is in equilibrium with the drain, the 2DEG beyond $G_2$. Thus, any voltage difference induced between the source and the drain drops on the tunnel junction between $G_1$ and $G_2$.

\begin{figure}
\centering\includegraphics[width=10cm,angle=0,clip=]{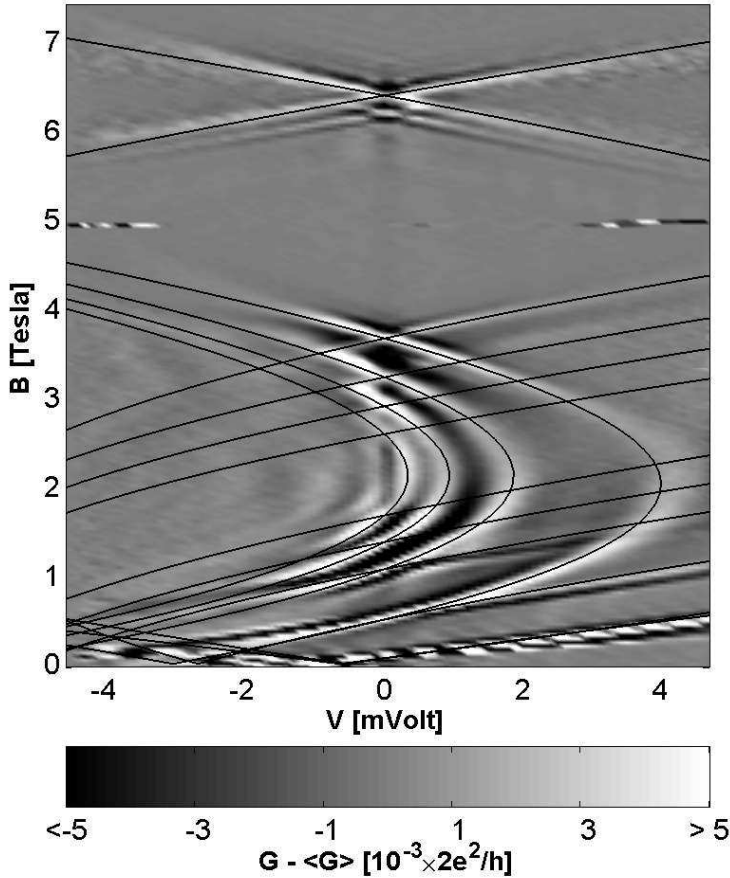}
\caption{\label{GVB}Plot of \GVB\ for a 10\micro{m}\
junction. In order to highlight the features a smoothed background has been subtracted from the raw data. The dispersions of the upper wire are easily discernable. Also plotted are fits to the free electron model with $m^*=0.75m_{\text{GaAs}}$ for the lower wire and $m^*=0.85m_{\text{GaAs}}$ for the upper wire (corresponding to $g_{\text{U}}=0.85$ and $g_{\text{L}}=0.75$). The extracted (zero field) densities are: $\{n_{\text{L}}\}=88,47,32,24,15\pm1\mu\mbox{m}^{-1}$ for the lower wire and 
$\{n_{\text{U}}\}=90,64,62\pm1\mu\mbox{m}^{-1}$ for the upper wire.}
\end{figure}
A typical measurement of \GVB\ for a long junction is shown in \fig{GVB}. The most prominent features in the figure are parabolic-like curves. In the noninteracting electron picture, these curves would mark boundaries in the $\vsd-B$ plane between regions where tunneling is blocked and regions where it is allowed. As explained in \cite{Ophir02}, each of these curves is the dispersion of a mode in one of the wires. With the voltage convention in \fig{circuit}, curves with turning points at $\vsd>0$ are lower wire dispersions and those with turning points at $\vsd<0$ are upper wire dispersions. At $\vsd=0$ only electrons at the Fermi level can participate in tunneling. Therefore tunneling is appreciable only if the momentum transferred to a tunneling electron is given by $\pm\left(k^U_F\pm k^L_F\right)$, where $k^{U,L}_F=\pi n_{\text{U,L}}/2$ in a spin-degenerate mode ($n_{\text{U,L}}$ denote electron densities of modes in the upper and lower wires and Zeeman splitting is ignored because it is not important here). This happens only when the magnitude of the magnetic field is:
\begin{equation}\label{xpnts}
    B^\pm=\frac{\hbar}{ed}\left|k^U_F\pm k^L_F\right|.
\end{equation}
Thus $B^\pm$ are a direct measure of the sum and difference of the densities of each of the modes in each of the wires. This allows to calculate the dispersions non-interacting electrons would have at the same density, because they depend only on the band mass $m_{\text{GaAs}}$ and the parameters of the heterostructure, which are known. We find that the observed dispersions deviate significantly from the non-interacting dispersions. 

\begin{figure}
\includegraphics[width=15cm,angle=0,clip=]{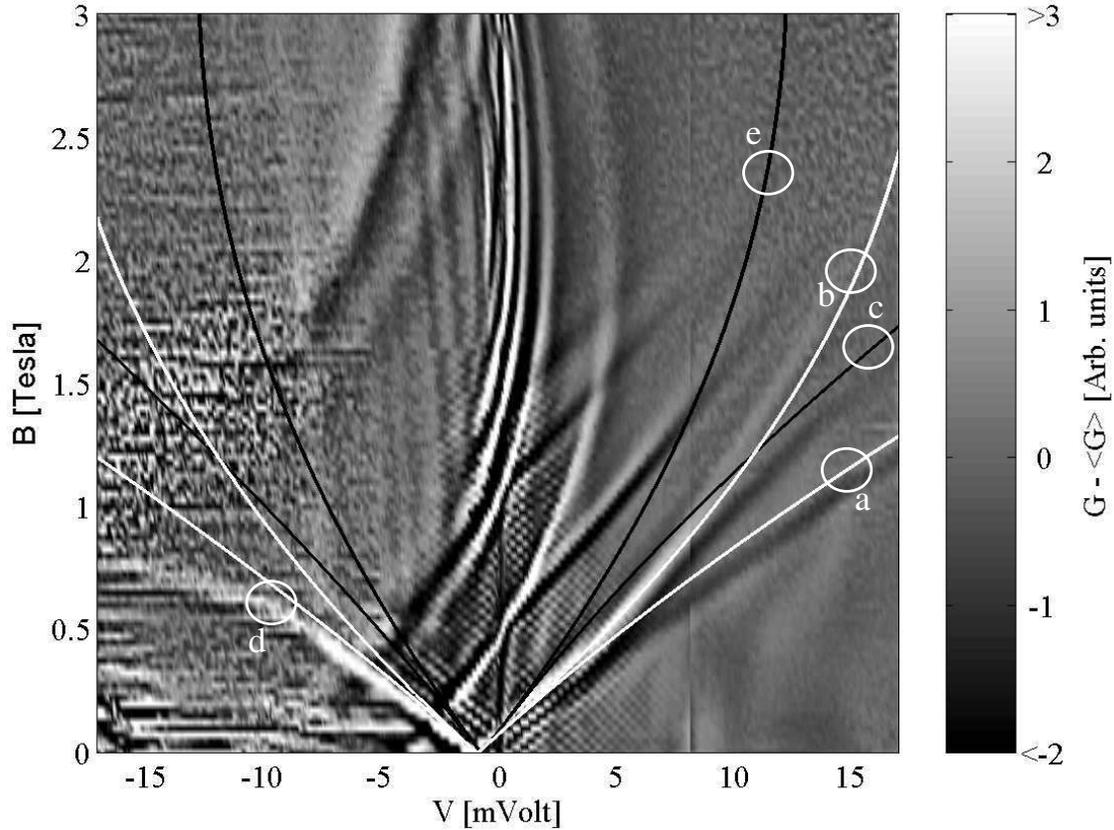}
\caption{\label{GVB_lowB}High resolution plot of \GVB\ for a 2\micro{m}\ junction. In order accentuate weak features a smoothed background has been subtracted, and the color-scale made nonlinear. Light shows positive and dark negative signal. The black curves are the expected dispersions of noninteracting electrons at the same electron density as the lowest-energy 1D bands of the wires, as determined from the crossing points with the $B$-axis (only the lower one is shown in the figure). The white curves are generated in a similar way but with a renormalized GaAs band-structure mass: $m^*=0.7m_{\text{GaAs}}$. This corresponds to $g_{\text{U}}=g_{\text{L}}=0.7$. Only the curves labelled a,b,c \& d in the plot are found to trace out experimentally-observable peaks in \GVB\ with the curve d following the measured peak only at $V>-10$mV.}
\end{figure}

For quantitative comparison with non-interacting dispersions one has to account for the fact that in finite width wires, like those reported here, the dispersion of each mode depends on magnetic field. As $B$ is increased the dispersions become more flat (when $B$ is extremely large Landau levels are be recovered). This affects the densities, the general trend being that lower-energy modes are populated at the expense of higher-energy modes. Solving the Schr\"{o}dinger equation numerically for a finite square potential well in the growth direction, we determine the zero field occupations from the $B^\pm$'s and find the shape of the curves traced out in the $\vsd-B$ plane by the dispersions. When this is done we find a systematic discrepancy - the real traces have an enhanced curvature. To model this observation we solve the finite well problem again, but with a renormalized mass, $m^*=gm_{\text{GaAs}}$, where $0<g<1$. This corresponds to a dispersion with the same density as for $m_{\text{GaAs}}$, that has an enhanced curvature and hence an enhanced Fermi-velocity, $v_F/g$. In all of the cases we have studied, $g$ is significantly suppressed from 1. For example, in \fig{GVB} we find for the lower wire $g_{\text{U}}=0.75$ and for the upper wire $g_{\text{U}}=0.85$, whereas in \fig{GVB_lowB} we find $g_{\text{L}}=g_{\text{U}}=0.7$. While the understanding of the curvature enhancement is still lacking, the enhanced velocity is consistent with Luttinger-liquid theory. The theory predicts low voltage peaks in the tunneling conductance that trace out dispersions of charge modes that have a velocity $v_c=v_F/g$ \cite{Carpentier02}.

The high resolution scan of a 2\micro{m} junction shown in \fig{GVB_lowB} reveals interesting extra features. Zooming on the low field region the peaks tracing the dispersions can be seen clearly. We have overlayed the plot with our simple calculation of the dispersions: In black we plot the dispersions with $m^*=m_{\text{GaAs}}$ while in white we plot the dispersions obtained with $m^*=0.7m_{\text{GaAs}}$. This value of $m^*$ was chosen to fit the observed dispersions also near $B^+$ (not shown in the figure). Lines a,b \& d describe the dispersions rather well, at least above $-10$mV, leading us to conclude that both wires possess modes with $v_c=v_F/0.7$. Interestingly there is an extra mode in the upper wire, marked c that lies very close to a noninteracting curve belonging to the upper wire. We can thus conclude that there is a mode in the upper wire that moves at approximately $v_F$, as expected in Luttinger-liquid theory for a spin mode.

\begin{figure}[ht]
\begin{center}
\centering\includegraphics[width=7.5cm,angle=0,clip=]{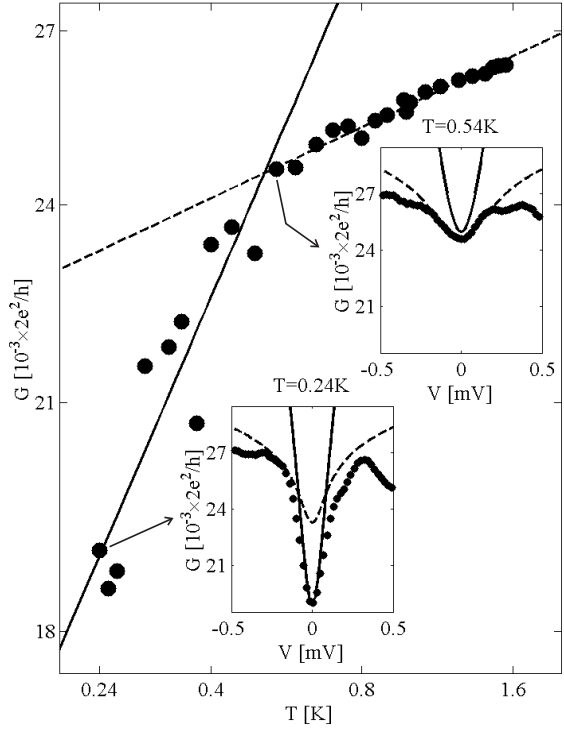}
\end{center}
\caption{\label{VTdep}Conductance near $\vsd=0$ as a function of temperature at $B=2.5\mbox{T}$ ($\bullet$) for a 6\micro{m}\ junction. The data was extracted from scans such as the those shown in the insets. Also shown are the fits to $T^{\alpha_{\text{end}}(g)}$ (solid line) and $T^{\alpha_{\text{bulk}}(g)}$ (dashed line), where we used $g=0.59$.
{\bf Insets:} Non-linear tunneling conductance at $B=2.5$T as a function of $\vsd$ for $T=0.24$K and $T=0.54$K. Here we show fits to \eq{leon} with $\alpha(T)=\alpha_{\text{end}}$ (solid line) and $\alpha(T)=\alpha_{\text{bulk}}$ (dashed line).}
\end{figure}

In addition to the dispersions seen in \fig{GVB_lowB} this figure contains features that we attribute to the finite length of the junction. The first of these is an intricate pattern of oscillations, to which we return below. The second is a zero bias anomaly that is seen as a sharp crease near $\vsd=0$. \fig{VTdep} shows that this dip is very sensitive to temperature. 

The source of the dip is the suppression of the tunneling density of states characteristic of a Luttinger liquid \cite{Tserkovnyak03}. For the range of $B$ in \fig{GVB_lowB}, the signal is dominated by tunneling from the Fermi-liquid 2DEG (having a constant density of states at the Fermi level) in the upper well to a Luttinger-liquid in the lower well. The zero-bias dip is thus entirely due to electron correlations in the lower wire. This is manifested by \cite{Bockrath99}:
\begin{equation}\label{leon}
  G(\vsd,T)\propto T^{\alpha(T)}F_{\alpha(T)}\left(\frac{e\vsd}{k_B T}\right),
\end{equation}
where $F_\alpha$ is a known scaling function obeying $F_\alpha(x)\sim1$ as $x\rightarrow0$ and $F_\alpha(x)\sim x^\alpha$ for $x\gg1$. The exponent $\alpha$ obeys \cite{Tserkovnyak03}:
\begin{eqnarray}\label{alpha}
  \alpha(T)\sim
  \left\{\begin{array}{cc}
\alpha_{\text{bulk}} &~~k_BT\gg \hbar v_F/(gL)\\
\alpha_{\text{end}}  &~~k_BT\ll \hbar v_F/(gL).\\
        \end{array}\right.
\end{eqnarray}
$\alpha_{\text{bulk,end}}$ are the exponents obtained for tunneling into the middle and into the end of a Luttinger liquid \cite{Kane97}:
\begin{eqnarray}
\alpha_{\text{bulk}}&=&(g+g^{-1}-2)/4, \label{alpha_b}\\
\alpha_{\text{end}}&=&(g^{-1}-1)/2 \label{alpha_e}.
\end{eqnarray}
In \cite{Tserkovnyak03} it is argued that $\alpha_{\text{bulk}}$ occurs because when $T$ is not too small (relative to $v_F/(gL)$), the tunneling process is insensitive to its occurring near the end of the lower wire, so the exponent $\alpha$ is characteristic of tunneling into the middle of a Luttinger liquid (\ref{alpha_b}). For lower values of $T$ the tunneling is effectively into the end of the lower wire so the exponent is the exponent for tunneling into the end of a Luttinger liquid (\ref{alpha_e}).

To compare the data in \fig{VTdep} to \eq{leon} we first fit the data to the $\vsd=0$ limit of \eq{leon}:
\begin{equation}\label{powerlaw}
  G\sim T^{\alpha(T)},
\end{equation}
where $\alpha(T)$ is given by \eq{alpha}. The result of the fit is overlayed on the data in \fig{VTdep}. As a corroboration we use the parameters from the fit in \eq{leon} with either $\alpha_{\text{bulk}}$ or $\alpha_{\text{end}}$. The results are plotted in the insets of \fig{VTdep}, where one can see that at low $T$ and \Vsd\ the data is reasonably well described by \eq{leon} with $\alpha(T)=\alpha_{\text{end}}$ while when $T$ or \Vsd\ increase there is a crossover and $\alpha(T)=\alpha_{\text{bulk}}$.

\begin{figure}[ht]
\centering\includegraphics[width=12cm,angle=0,clip=]{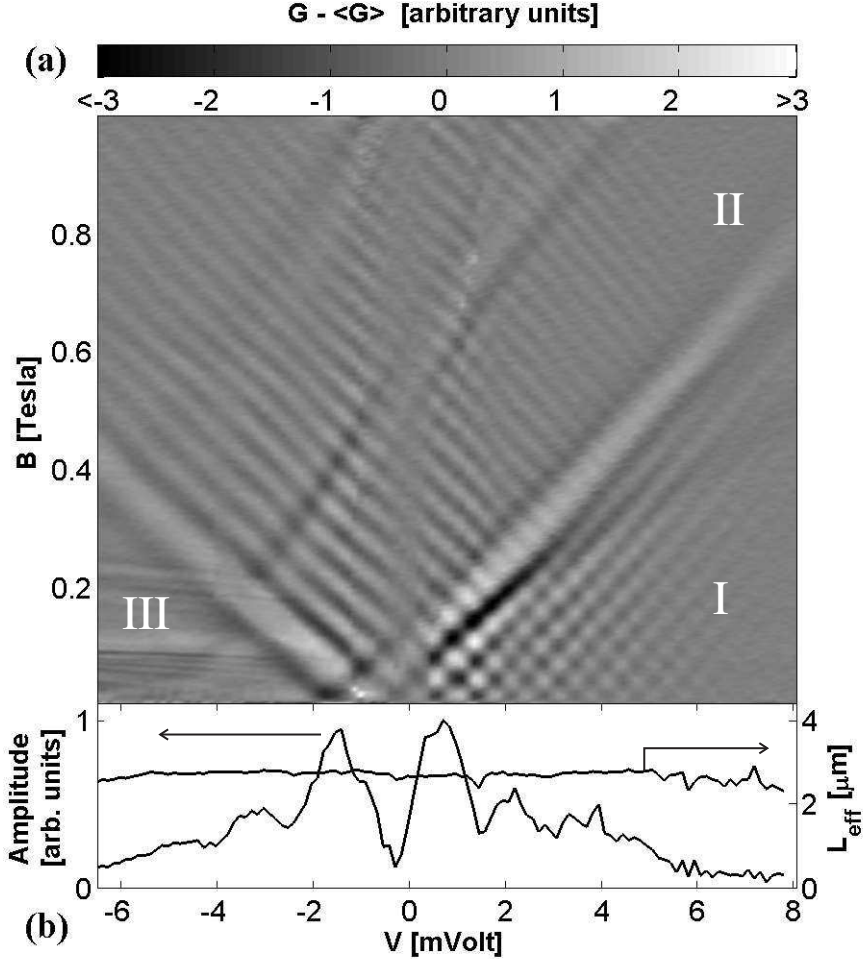}
\caption{{\bf (a)} Conductance oscillations at low field from a 2\micro{m} junction.
A smoothed background has been removed from the raw data and the nonlinear scale has been optimized to increase the visibility of the oscillations. Parallel side lobes attributed to finite-size effects appear only to the right of the main dispersion peaks, defining quadrants I, II and III. Also present is a slow modulation of the interference along the \Vsd-axis. {\bf (b)} Absolute value and position (converted to length) of the peak corresponding to the oscillations along $B$ in quadrant II of (a), as determined from the Fourier transform of $S^{1-1/\beta}G\left(V,S^{1+1/\beta}\right)$. See main text for definition of $S$, $\beta$ and other details. The slow modulation as a function of \Vsd\ is easily discerned.}\label{2microns}%
\end{figure}
\begin{figure}[ht]
\centering\includegraphics[width=12cm,angle=0,clip=]{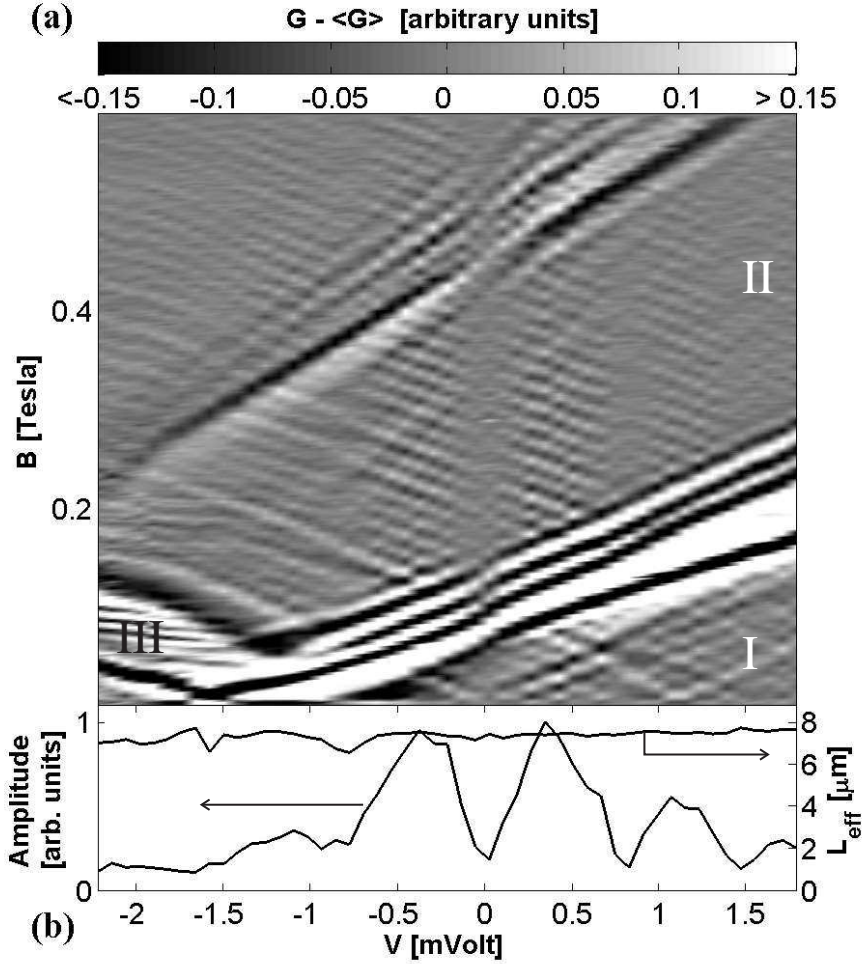}
\caption{Same as \fig{2microns} but for a 6\micro{m} junction. Note that the oscillations are approximately three times faster than in \fig{2microns}, as expected from \eq{AB}. For this junction, the asymmetry in the strength of the side lobes on opposite sides of a dispersion peak is less pronounced than for the shorter junction appearing in \fig{2microns}.}\label{6microns}%
\end{figure}

We now return to the oscillation pattern seen in \fig{GVB_lowB} and in \figs{2microns}, \ref{6microns}. In the last two figures the range of field is such that the lines that correspond to the dispersion curves appear as pronounced peaks that extend diagonally across the figures. In addition to these we observe numerous secondary peaks running parallel to the dispersions of the lowest-energy modes. These side lobes always appear to the right of upper wire dispersions, in the region that corresponds to momentum conserving tunneling for an upper wire with a reduced density. Thus, the $V-B$ plane is separated into quadrants: quadrant I has a checkerboard pattern of oscillations, quadrant II has a hatched pattern and quadrant III has no regular pattern. Furthermore, the frequency of the oscillations depends on the length of the junction. When the lithographic length of the junction, $L$, is increased from 2\micro{m} (\fig{2microns}) to 6\micro{m} (\fig{6microns}) the frequency in bias and in field increases by a factor of approximately three. The period is related to the length of the junction, $L$, by the formula:
\begin{equation}\label{AB}
  \Delta \vsd L/v_p=\Delta BLd\approx\phi_0,%
\end{equation}
where $\phi_0=h/e$ is the quantum of flux and $v_p$ is an effective velocity. \eq{AB} can be used to extract $L$, because $d$ is known.

To understand the asymmetry of the interference around the dispersion curves it suffices to consider non-interacting electrons in both wires. Then, for weak tunneling, the current is related by the Fermi golden rule to the tunneling matrix element between a state in one wire and a state in the other \cite{Tserkovnyak02,Tserkovnyak03}:%
\begin{equation}\label{nonint}
I(\vsd,B)\sim \vsd\left[\left|M\left(\kappa_+\right)\right|^2+\left|M\left(\kappa_-\right)\right|^2\right],
\end{equation}
where
\begin{equation}\label{kappa}
    M\left(\kappa\right)=\int_{-\infty}^{\infty} dx e^{i\kappa x}\psi_U(x)e^{-ik^U_F x},
\end{equation}
and $\kappa_\pm=k^U_F-k^L_F+e\vsd/(\hbar v_F)\pm q_B$. Here $v_F$ is the Fermi velocity, which is almost the same in both modes giving rise to the interference. We shall further assume that the potential limiting the upper wire's length, $U(x)$, is smooth on the scale of the Fermi wavelength, a reasonable assumption since it is defined by top gates lying 0.5\micro{m} away. Under this assumption the WKB approximation can be used to write:
\begin{equation}\label{WKB}
    \psi_U(x)\sim k^{-1/2}(x)\exp\left[ik_{F}^Ux-is(x)\right],
\end{equation} 
where $k(x)=k_{F}^U\sqrt{1-U(x)/E^U_F}$ and $s(x)=\int_0^xdx'\left[k^U_F-k(x')\right]$. \eq{nonint} can be used to find $U(x)$. We have found that to a good approximation it is given by $U(x)\approx E^U_F\left|2x/L_{\text{eff}}\right|^\beta$. In this model, the function $S^{1-1/\beta}G\left(\vsd,S^{1+1/\beta}\right)$ is periodic in $S^{1+1/\beta}$, with a period determined by $L_{\text{eff}}$, and where $S=\hbar\kappa_+/(ed)$ is in quadrant II in \figs{2microns}a and \ref{6microns}a. We find that the data is well described by $\beta=8\pm2$ for the short junction and  $\beta=21.5\pm2$ for the long junction \cite{Tserkovnyak03}. With these values of $\beta$ we then perform Fourier analysis of the data for each value of \Vsd. The results for the data in \fig{2microns}a are plotted in \fig{2microns}b, were both the amplitude of the main peak and its position are shown. In the figure we convert the frequency of the oscillations back to the effective length for the upper wire, $L_{\text{eff}}(\vsd)$, which clearly depends only very weakly on \Vsd. Similar analysis was performed for the data in \fig{6microns}a. The length extracted from the Fourier analysis for both the 2\micro{m} junction ($L_{\text{eff}}=2.7\pm0.1$\micro{m}) and for the 6\micro{m} junction ($L_{\text{eff}}=7.3\pm0.3$\micro{m}) is larger than the lithographic length. This is reasonable because the gates delimiting the upper wire are on the surface of the sample.

Next we turn our attention to the structure of the amplitude of the main Fourier peak (see panels b in \figs{2microns}, \ref{6microns}). The amplitude is seen to oscillate as a function of \Vsd, giving rise to a series of vertical strips of suppressed conductance in \figs{2microns}a and \ref{6microns}a. To understand this effect the finite interactions in the wires have to be taken into account. A more general expression for the tunneling current (to lowest order in tunneling) is given by \cite{Carpentier02}:
\begin{equation}\label{I_int}
I(\vsd,B)\propto\int_{-\infty}^{\infty}dx\int_{-\infty}^{\infty}dx'\int_{-\infty}^{\infty}dt
e^{iq_B(x-x')}e^{ie\vsd t/\hbar}C(x,x';t).
\end{equation}
where $C(x,x';t)$ is the two point Green function, which is known from Luttinger liquid theory. In the case we are describing here one finds that the interference has a contribution from two velocities in the upper wire, $v_F$ and a charge mode velocity, $v^-_c$. $v^-_c$ is the velocity of the antisymmetric charge mode which arises because the Fermi velocities in the two wires are very similar. As a result of this, the interference in \figs{2microns} and \ref{6microns} can be understood as a moir\'{e} pattern created by side lobes running with a slope (in the $\vsd-B$ plane) $(v_Fd)^{-1}$ and other side lobes with a slope $(v^-_cd)^{-1}$. From the ratio between $\Delta V$ (defined in \eq{AB}) and the distance between the suppression strips, $\Delta V_{\text{mod}}$, the ratio between the two velocities can be extracted. One finds: 
\begin{equation}\label{moire}
    \frac{\Delta V_{\text{mod}}}{\Delta V}=\frac12\frac{1+g_-}{1-g_-},
\end{equation}
where $g_-=v_F/v^-_c$. For the data presented here we find $g_-=0.67\pm0.07$. This is in agreement with previous assessments of $g$ in CEO wires and is a direct consequence of spin-charge separation in our wires.

This work was supported in part by the US-Israel BSF, the European Commission RTN
Network Contract No. HPRN-CT-2000-00125 and NSF Grant DMR 02-33773. YT is supported by the Harvard Society of Fellows.

\newcommand{\noopsort}[1]{} \newcommand{\printfirst}[2]{#1}
  \newcommand{\singleletter}[1]{#1} \newcommand{\switchargs}[2]{#2#1}

\end{document}